# Features of Magneto-Optical Resonances in an Elliptically Polarized Traveling Light Wave

D. V. Brazhnikov[a,b], A. V. Taĭchenachev[a,c], A. M. Tumaĭkin[a], V. I. Yudin[a,b], S. A. Zibrov[d,e], Ya. O. Dudin[d,e], V. V. Vasil'ev[d], and V. L. Velichansky[d]

[a] *Institute of Laser Physics, Siberian Division, Russian Academy of Sciences, pr. Akademika Lavrent'eva 13/3, Novosibirsk, 630090 Russia*
e-mail: llf@laser.nsc.ru
[b] *Novosibirsk State Technical University, ul. Pirogova 2, Novosibirsk, 630092 Russia*
[c] *Novosibirsk State University, ul. Pirogova 2, Novosibirsk, 630090 Russia*
[d] *Lebedev Physical Institute, Russian Academy of Sciences, Leninskiĭ pr. 53, Moscow, 117924 Russia*
[e] *Moscow Engineering Physics Institute (State University), Kashirskoe sh. 31, Moscow, 115409 Russia*

Received December 8, 2005

The parameters of nonlinear absorption magneto-optical resonances in the Hanle configuration have been studied as functions of the ellipticity of a traveling light wave. It has been found that these parameters (amplitude, width, and amplitude-to-width ratio) depend strongly on the polarization of the light wave. In particular, the resonance amplitude can increase by more than an order of magnitude when the polarization changes from linear to optimal elliptic. It has been shown that this effect is associated with the Doppler frequency shift for atoms in a gas. The theoretical results have been corroborated in experiments in Rb vapor.

PACS numbers: 33.55.Fi, 42.50.Gy, 42.62.Fi

**1.** Nonlinear interference effects based on atomic coherence attract special continuously increasing interest in recent time. These effects are widely applied in ultra-high-resolution nonlinear spectroscopy [1, 2], metrology (atomic clocks and magnetometers) [3], laser cooling [4], and other modern areas of laser physics.

The typical induced-absorption resonance has the shape of a narrow dip (electromagnetically induced transparency, EIT) or peak (electromagnetically induced absorption, EIA). The first type of the resonance has long been known (see, e.g., [5, 6]) and thereby is better studied than the EIA resonance, which was observed in 1998 in two-frequency excitation [7] and then in the Hanle configuration [8]. As was shown in [9, 10], EIA is physically caused by the transfer of anisotropy (Zeeman coherence) from the excited level to the ground one. The possibility of change from EIA to EIT was analyzed in [11] for the Hanle configuration in dependence on depolarizing relaxation in the exited state.

In this work, the parameters (amplitude, width, and amplitude-to-width ratio) of the magneto-optical resonance are studied as functions of the ellipticity of the light wave, which resonantly excites the closed optical transition $F_g \longrightarrow F_e$ in the atomic gas, where $F_g$ and $F_e$ are the total angular momenta of the atom in the ground and excited states, respectively. The total absorption of the traveling light wave at a certain magnetic field is a spectroscopic signal. In the case under consideration, the magnetic field is directed along the wave vector of the light wave (Hanle configuration). In contrast to previous works (e.g., [8, 12, 13]), where the case of linear polarization was considered, we analyze the general case of the elliptic polarization of the electromagnetic wave with emphasis on the EIA resonance. Certain new features attributed to the Doppler effect for atoms in a gas are theoretically revealed. These features are experimentally corroborated for the $F_g = 2 \longrightarrow F_e = 3$ transition in the $D_2$ line of the $^{87}$Rb atom.

**2.** We consider the resonance interaction of the elliptically polarized plane traveling light wave

$$\mathbf{E}(z, t) = E_0 \mathbf{e} \exp\{-i(\omega t - kz)\} + \text{c.c.}$$

with atoms where the closed optical transition $F_g \longrightarrow F_e$ occurs between the ground and excited states. Here, $\omega$ and $E_0$ are the frequency and amplitude of the light field and the unit complex elliptic-polarization vector has the form

$$\mathbf{e} = \mathbf{e}_x \cos(\varepsilon) + i\mathbf{e}_y \sin(\varepsilon)$$
$$= -\mathbf{e}_{+1}\cos(\varepsilon - \pi/4) - \mathbf{e}_{-1}\sin(\varepsilon - \pi/4),$$

where $\mathbf{e}_{\pm 1} = \mp(\mathbf{e}_x \pm i\mathbf{e}_y)/\sqrt{2}$ are the cyclic basis vectors. The ellipticity parameter $\varepsilon$ is defined in the interval $-\pi/4 \leq \varepsilon \leq \pi/4$ so that $|\tan(\varepsilon)|$ is the ratio of the small semiaxis of the ellipse to its large semiaxis and the sign of this parameter specifies the rotation direction for the instantaneous strength of the light field (see Fig. 1a).

The magnetic field has the form $\mathbf{B} = B\mathbf{e}_0$, where $\mathbf{e}_0$ is the unit vector directed along $\mathbf{k}$. The Hamiltonian of the interaction of an atom with the external field is $\hat{H} = \hat{H}_B + \hat{H}_E$, where the Hamiltonian $\hat{H}_B$ of the interaction of the atom with magnetic field and the Hamiltonian $\hat{H}_E$ of the interaction of the atom with the traveling electromagnetic wave have the form

$$\hat{H}_E = \hbar R \exp\{-i(\omega t - kz)\}\hat{V} + \text{H.c.}, \quad (1)$$

$$\hat{H}_B = \sum_{a=e,g} \hbar\Omega_a \hat{F}^a. \quad (2)$$

Here, $R = -dE_0/\hbar$ is the Rabi frequency ($d$ is the reduced matrix element of the atomic dipole moment), $\Omega_a = \mu_B g_a B_0/\hbar$ corresponds to the Zeeman sublevel splitting ($g_a$ is the Landé factor and $\mu_B$ is the Bohr magneton), $\hat{F}^a$ are the operators of the total angular momenta of the levels, and the operator $\hat{V}$ is expressed in terms of the Clebsch–Gordan coefficients as

$$\hat{V} = \sum_{q=\pm 1, m_e, m_g} C^{F_e, m_e}_{F_g, m_g; 1q} e^q |F_e, m_e\rangle\langle m_g, F_g|,$$

where $e^{+1} = -\cos(\varepsilon - \pi/4)$ and $e^{-1} = -\sin(\varepsilon - \pi/4)$ are the components of the unit complex elliptic polarization vector in the cyclic basis. The generalized optical Bloch equations for the stationary density matrix in the rotating basis have the form [11]

$$(\gamma_{eg} - i\delta)\hat{\rho}^{eg} = -iR(\hat{V}\hat{\rho}^g - \hat{\rho}^e\hat{V}) \\ - i(\Omega_e \hat{F}^e \hat{\rho}^{eg} - \Omega_g \hat{\rho}^{eg}\hat{F}^g), \quad (3)$$

$$(\Gamma + \gamma_r)\hat{\rho}^e = -iR(\hat{V}\hat{\rho}^{ge} - \hat{\rho}^{eg}\hat{V}^+) - i\Omega_e[\hat{F}^e, \hat{\rho}^e], \quad (4)$$

$$\Gamma[\hat{\rho}^g - \hat{\rho}^g_0] = \hat{\gamma}_r\{\hat{\rho}^e\} - iR(\hat{V}^+\hat{\rho}^{eg} - \hat{\rho}^{ge}\hat{V}) \\ - i\Omega_g[\hat{F}^g, \hat{\rho}^g]. \quad (5)$$

Here, the density matrices $\hat{\rho}^g$ and $\hat{\rho}^e$ describe the distributions over the magnetic sublevels of the ground and excited states, respectively, and the matrices $\hat{\rho}^{ge} = \hat{\rho}^{eg\dagger}$ describe the optical coherences (in the rotating basis). The isotropic distribution of $\hat{\rho}^g_0$ over the magnetic sublevels of the ground state in the absence of the light field is normalized by the condition $\text{Tr}\{\hat{\rho}^g_0\} = 1$. The detuning of the light-field frequency from the transition frequency $\omega_0$ is equal to $\delta = \omega - \omega_0 - kv$, where $kv$ is the Doppler shift at the atomic velocity projection $v$ onto the $z$ axis. The relaxation processes are described by the following relaxation constants: $\gamma_r$ is the radiative relaxation rate for the excited state, the constant $\Gamma$ specifies the in-flight relaxation, $\gamma_{eg} = \gamma_r/2 + \Gamma$ is the relaxation rate for optical coherences, and the operator $\hat{\gamma}_r\{\hat{\rho}^e\}$ describes the transition of the atoms from the excited level to the ground one due to spontaneous relaxation. Collisional depolarization is disregarded in this work.

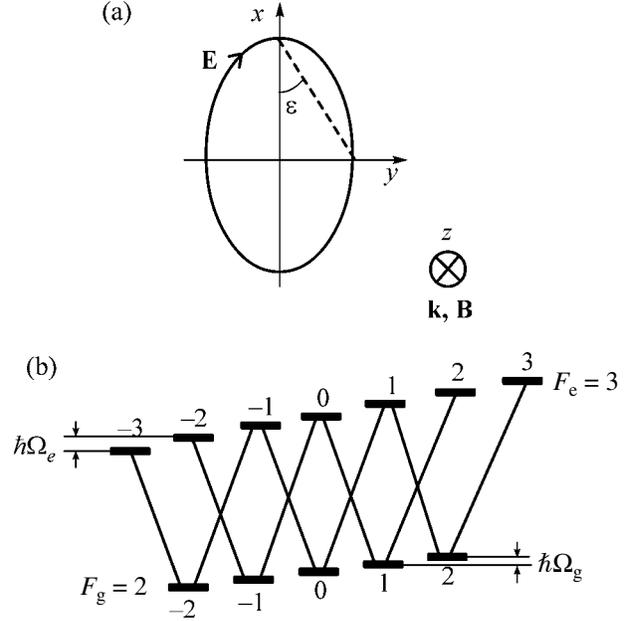

**Fig. 1.** (a) Mutual orientation of the polarization ellipse, the wave vector of the light wave, and the magnetic field; (b) the scheme of the electromagnetic induced transitions.

**3.** The system of Eqs. (3)–(5) was solved numerically. The signal of the complete absorption of the traveling light wave in an optically thin medium was analyzed with allowance for the Maxwellian velocity distribution. This signal has a rather complicated shape and consists of structures with various widths (see Fig. 2). We focus on the narrowest central structure. The calculations were performed for the exact one-photon resonance $\omega = \omega_0$.

Special attention was paid to the EIA resonance on the closed optical transition $F_g = 2 \longrightarrow F_e = 3$ (Fig. 1b). For example, Fig. 3 shows two magneto-optical EIA resonances for linear ($\varepsilon = 0$) and elliptic ($\varepsilon = \pi/9$) polarizations of the light wave (the Rabi frequency $R = 5\gamma_{eg}$, the in-flight relaxation constant $\Gamma = 5 \times 10^{-3}\gamma_{eg}$, and the spontaneous relaxation rate $\gamma_r = 2\gamma_{eg}$). As is seen in Fig. 3, the parameters of the nonlinear resonances depend strongly on the polarization of the traveling wave.

**4.** In order to verify the theoretical results, the corresponding experiment was carried out. Figure 4 presents the layout of the experimental setup and the scheme of the atomic levels. Resonance radiation is generated by an injection laser designed according to the Litrov scheme (ECDL). The design of the laser allows the smooth variation of the radiation frequency over a wide range (up to 40 GHz). To study the characteristics of the electromagnetic induced absorption resonance, the laser is tuned to the optical transition $5^2S_{1/2}$, $F = 2 \longrightarrow 5^2P_{3/2}$, $F' = 3$ (the $D_2$ line, λ = 780 nm). A small fraction of the laser radiation is separated by a splitter (BS) and is directed to the frequency control system. The scheme and operating grounds of the control system were reported in [14]. After the splitter, the laser radiation sequentially passes through a half-wave plate (λ/2) and crystal polarizer (Polarizer), which allows the smooth variation of the radiation power. The ellipticity of the resonance radiation is smoothly varied by means of a quarter-wave plate (λ/4) located immediately after the polarizer.

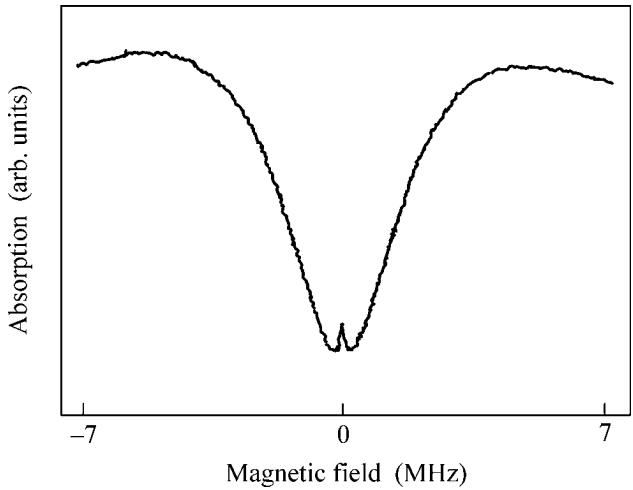

**Fig. 2.** Magneto-optical EIA resonance for a beam with a diameter of 20 mm, a power of 1.9 mW, and ε = 0.

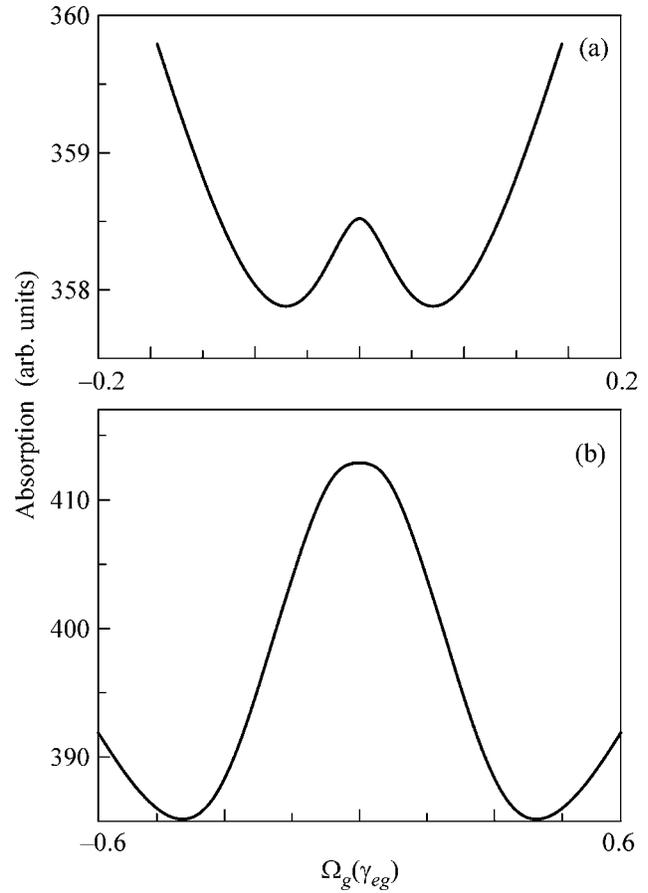

**Fig. 3.** Magneto-optical EIA resonance for (a) linear polarization (ε = 0), $A = 0.65$, $W = 0.06\gamma_{eg}$, and $A/W = 10.8\gamma_{eg}^{-1}$ and (b) elliptic polarization (ε = π/9), $A = 27.74$, $W = 0.4\gamma_{eg}$, and $A/W = 69.35\gamma_{eg}^{-1}$.

The experiment was carried out with a cylindrical glass cell (60 mm in length and 55 mm in diameter) containing isotopically pure atomic $^{87}$Rb vapor. The cell was placed in a solenoid, which ensured the variation of the longitudinal magnetic field. The splitting of the magnetic sublevels in $^{87}$Rb is equal to $B \times 0.7$ MHz/G. For the isolation from the external laboratory field, the cell was placed in three concentric cylindrical magnetic screens. The temperature of the cell was equal to 20°C. The laser radiation power at the entry to the cell was equal to 3 mW and the beam diameter was 5 mm. The intensity of the radiation passed through the cell was detected as a function of the longitudinal magnetic field by a photodiode whose signal was observed on an oscilloscope screen.

**5.** Figure 5 shows the calculated and measured parameters (amplitude, width, and amplitude-to-width ratio) for the EIA resonance as functions of the ellipticity ε for the $F_g = 2 \longrightarrow F_e = 3$ transition in the $D_2$ line of the $^{87}$Rb atom. According to this figure, the theory and experiment are in good qualitative agreement with each other. The dependences exhibit a number of fundamental features for elliptic polarization. The amplitude and amplitude-to-width ratio as functions of the ellipticity have extrema at a certain elliptic polarization $\varepsilon_{max} \neq 0$. At the same time, according to the calculations, the maximum for immovable atoms corresponds to linear polarization, i.e., $\varepsilon_{max} = 0$. Thus, the situation changes dramatically for the thermalized atomic gas. This effect is most pronounced in quite strong light fields when the central-resonance amplitude at the extremum point $\varepsilon_{max}$ is one or two orders of magnitude larger than the resonance amplitude for linear polarization ε = 0 (by a factor of 40 for Fig. 5a). The results obtained above show that the presence of the maximum in the amplitude of the EIA resonance at a certain elliptic polarization $\varepsilon_{max} \neq 0$ is attributed to the Doppler frequency shift and the motion of atoms in the gas. Indeed, different velocity groups of atoms have different one-photon detunings due to the Doppler effect. Therefore, as was shown in [11], the line contour for the case of elliptic polarization is deformed (becomes asymmetric and shifted). As a result, the line contour integrated (averaged) over all the velocity groups acquires qualitatively new features as compared to the case of immovable atoms. Some of such effects were previously described, e.g., the Doppler narrowing of the absorption line in the three-level Λ system [15]. The detection of the effect of the motion of atoms in the gas on the dependence of the amplitude of the EIA resonance on the ellipticity of the light field is the substantially new point of this work.

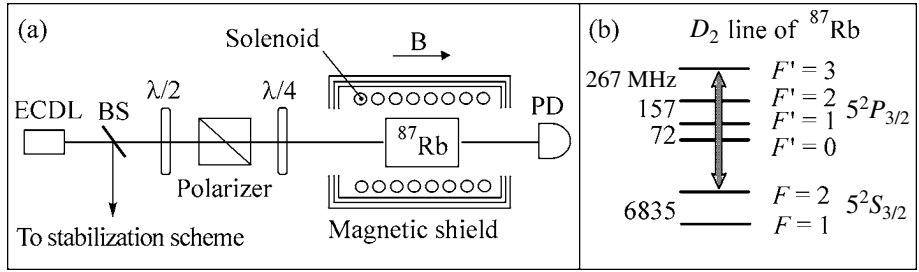

**Fig. 4.** (a) Layout of the experimental setup for the observation of the induced absorption resonance in the Hanle configuration: ECDL is the external-cavity diode laser, BS is the splitter, and PD is the photodetector; (b) the scheme of the atomic levels for the $D_2$ line of $^{87}$Rb.

It is worth noting that, in contrast to bright transitions, according to our calculations and experiments with $^{87}$Rb and $^{85}$Rb, the extremum in the ellipticity dependence of the amplitude of the EIT resonance for dark transitions ($F_g = 1 \longrightarrow F_e = 1$, $F_g = 1 \longrightarrow F_e = 0$, and $F_g = 2 \longrightarrow F_e = 1$) is observed at linear polarization $\varepsilon = 0$ similar to the case of immovable atoms.

6. The magneto-optical nonlinear EIA resonance has been investigated in this work both theoretically and experimentally for different polarizations (ellipticities $\varepsilon$) of the traveling light wave. A number of important results have been obtained. In particular, it has been found that the amplitude and amplitude-to-width ratio of the resonance can exhibit pronounced maxima at certain elliptic polarization ($\varepsilon \neq 0$) rather than at linear polarization ($\varepsilon = 0$). The amplitude of the nonlinear resonance at the optimum ellipticity $\varepsilon_{max}$ can be several tens of times larger than that for linear polarization. Thus, the parameters of the magneto-optical EIA resonance can be significantly (by one or two orders of magnitude) improved due to varying the ellipticity of the traveling light wave. The theoretical calculations have been qualitatively corroborated by the experimental data for the $D_2$ line of $^{87}$Rb: $5^2S_{1/2}$, $F = 2 \longrightarrow 5^2P_{3/2}$, $F' = 3$, $\lambda = 780$ nm. These results can be applied in some interesting areas such as the creation of the new generation of magnetometers. Some of the above results were reported at the conference ICONO/LAT-2005 [16].

This work was supported by the Russian Foundation for Basic Research (project nos. 04-02-16488, 05-02-17086, and 05-08-01389). The work of D.V.B. was sup-

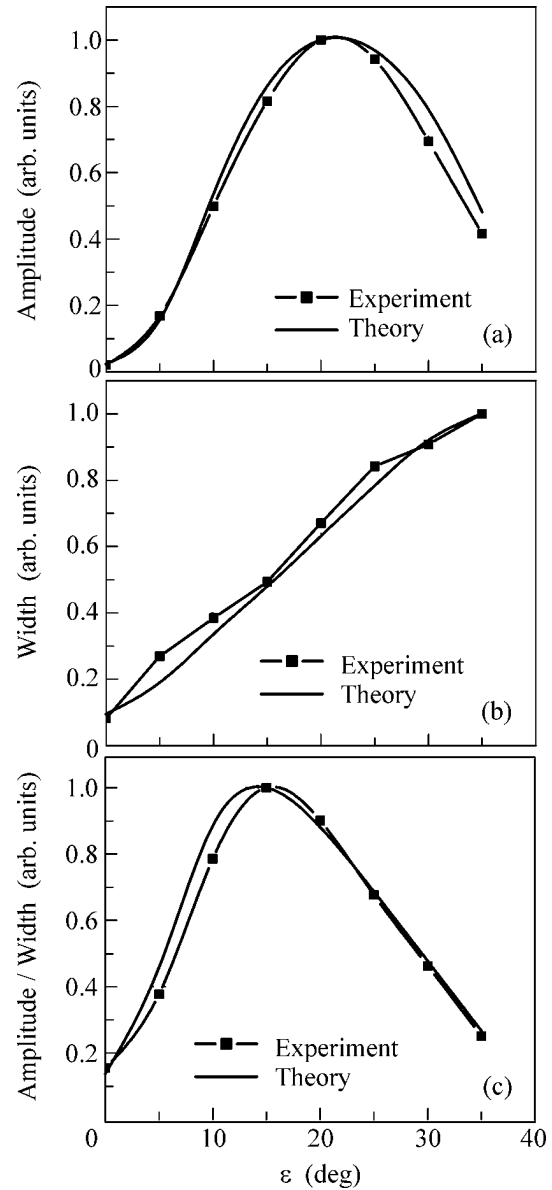

**Fig. 5.** (a) Amplitude, (b) width, and (c) amplitude-to-width ratio for the EIA resonance vs. the ellipticity of the polarization of the light wave.

## REFERENCES

1. M. Stahler, R. Wynands, S. Knappe, et al., Opt. Lett. **27**, 1472 (2002).
2. A. Akulshin, A. Celikov, and V. Velichansky, Opt. Commun. **84**, 139 (1991).
3. J. Vanier, Appl. Phys. B **81**, 421 (2005).
4. A. Aspect, E. Arimondo, R. Kaiser, et al., Phys. Rev. Lett. **61**, 826 (1988).
5. E. Arimondo, in *Progress in Optics*, Ed. by E. Wolf (Elsevier, Amsterdam, 1996), Vol. 35, p. 257.
6. B. D. Agap'ev, M. B. Gornyĭ, B. G. Matisov, and Yu. V. Rozhdestvenskiĭ, Usp. Fiz. Nauk **163** (9), 35 (1993) [Phys. Usp. **36**, 763 (1993)].
7. A. M. Akulshin, S. Barreiro, and A. Lezama, Phys. Rev. A **57**, 2996 (1998).
8. F. Renzoni, S. Cartaleva, G. Alzetta, and E. Arimondo, Phys. Rev. A **63**, 065401 (2001).
9. A. V. Taĭchenachev, A. M. Tumaĭkin, and V. I. Yudin, Pis'ma Zh. Éksp. Teor. Fiz. **69**, 776 (1999) [JETP Lett. **69**, 819 (1999)].
10. A. V. Taichenachev, A. M. Tumaikin, and V. I. Yudin, Phys. Rev. A **61**, 011802(R) (2000).
11. D. V. Brazhnikov, A. V. Taichenachev, A. M. Tumaikin, and V. I. Yudin, J. Opt. Soc. Am. B **22**, 57 (2005).
12. Y. Dancheva, G. Alzetta, S. Cartaleva, et al., Opt. Commun. **178**, 103 (2000).
13. F. Renzoni, C. Zimmermann, P. Verkerk, and E. Arimondo, J. Opt. B: Quantum Semiclassic. Opt. **3**, S7 (2001).
14. A. V. Yarovitskiĭ, O. N. Prudnikov, V. V. Vasil'ev, et al., Kvantovaya Élektron. (Moscow) **34**, 341 (2004).
15. A. V. Taĭchenachev, A. M. Tumaĭkin, and V. I. Yudin, Pis'ma Zh. Éksp. Teor. Fiz. **72**, 173 (2000) [JETP Lett. **72**, 119 (2000)].
16. D. V. Brazhnikov, A. V. Taichenachev, A. M. Tumaikin, et al., in *Technical Digest of ICONO/LAT-2005* (St. Petersburg, 2005), IFM19, Proc. SPIE (in press).

*Translated by R. Tyapaev*